# Perpendicular magnetic anisotropy, tunneling magnetoresistance and spin-transfer torque effect in magnetic tunnel junctions with Nb layers

Bowei Zhou,[1] Pravin Khanal,[1] Onri Jay Benally,[2] Deyuan Lyu,[2] Daniel B. Gopman,[3] Arthur Enriquez,[1] Ali Habiboglu,[1] Kennedy Warrilow,[1] Jian-Ping Wang,[2] and Wei-Gang Wang[1]*

1. Department of Physics, University of Arizona, Tucson, AZ 85721, USA
2. Department of Electrical & Computer Engineering, University of Minnesota, Minneapolis, MN 55455, USA
3. Materials Science & Engineering Division, National Institute of Standards and Technology, Gaithersburg, MD 20899, USA

Nb and its compounds are widely used in quantum computing due to their high superconducting transition temperatures and high critical fields. Devices that combine superconducting performance and spintronic non-volatility could deliver unique functionality. Here we report the study of magnetic tunnel junctions with Nb as the heavy metal layers. An interfacial perpendicular magnetic anisotropy energy density of 1.85 mJ/m$^2$ was obtained in Nb/CoFeB/MgO heterostructures. The tunneling magnetoresistance was evaluated in junctions with different thickness combinations and different annealing conditions. An optimized magnetoresistance of 120% was obtained at room temperature, with a damping parameter of 0.011 determined by ferromagnetic resonance. In addition, spin-transfer torque switching has also been successfully observed in these junctions with a quasistatic switching current density of $7.3 \times 10^5$ A/cm$^2$.

*wgwang@arizona.edu

Magnetic tunnel junctions (MTJs) [1–7] are critical components in magnetic random access memory (MRAM), [8,9] spin-logic, [10,11] spin-torque oscillator, [12–14] and neuromorphic [15] applications. Particularly, MTJs with perpendicular magnetic easy-axis (pMTJs) are promising due to their potential in ultrafast and ultralow energy operations,[16–19] and deep scaling capability down to a lateral size of 2-3 nm. [20] The perpendicular magnetic anisotropy (PMA) in many of these pMTJs has an interfacial origin, where the thickness of the ferromagnetic (FM) layer has to be maintained around 1 nm. [21–23] Typically the MgO barrier is sandwiched by two such thin FM layers, forming the core structure of the pMTJ. The other side of the thin FM layer is usually interfaced with a heavy metal (HM) layer. The perpendicular easy axis of the system is established due to the proper hybridization of the 3*d* wavefunction of the FM layer and the 2*p* wavefunction of the oxygen from the MgO barrier, which can be further influenced by the 3*d*-4 or $d$ 3$d$-5$d$ hybridization at the FM/HM interfaces.[21–23] Clearly in this case the magnetic and transport properties of pMTJs are sensitively dependent on the HM layers. A variety of HM layers such as Pt,[24] Ta,[16] Mo,[25,26] and W[22,27] have been explored to investigate the performance of pMTJs.

Niobium (Nb) and its compounds are one of the most important components in quantum computation due to their high superconductivity transition temperatures and high critical fields.[28] In addition to quantum computing, it is also interesting to study the behavior of pMTJs with Nb for cryogenic memory applications. Recently, it has been demonstrated that the magnetic anisotropy of Fe can be modified by the superconductivity of Vanadium in the MgO/Fe MTJs at low temperatures.[29] On the other hand, electrically modifying the magnetic properties of 3$d$ ferromagnets is important in many applications at room temperature (RT).[30–32] For example, through the voltage-controlled magnetic anisotropy (VCMA) effect,[31,33] a switching energy as low as a few femtojoules has been achieved in pMTJs.[16–19] Therefore it would be interesting to investigate the potential interaction between superconductivity and PMA, as well as the VCMA effect in Nb-based pMTJs. In the past, Nb has been successfully employed to obtain PMA in Nb/CoFeB/MgO heterostructures and the interfacial PMA energy density of 2.2 mJ/m$^2$ was obtained.[34–36] However, transport properties have not been evaluated in MTJs with Nb layers.

In this study, we report the RT performance of pMTJs with Nb as the HM layers. Magnetic, transport, and spin-dynamic properties of the junctions were investigated in blanket MTJ films as well as patterned junctions. The PMA and the thermal robustness of the pMTJs were studied by varying the thickness of the Nb layers and annealing conditions. A reasonably large TMR of 120 % was obtained at RT. Spin-transfer torque (STT)[37,38] switching has also been successfully observed in these junctions with a quasistatic switching current density of $7.3 \times 10^5$ A/cm$^2$.

The MTJ films in this work were fabricated in a 12-source UHV sputtering system (AJA International) with a base pressure of 10$^{-7}$ Pa (10$^{-9}$ Torr). The stack structure of the films is Si/SiO$_2$/Ta(8)/Ru(7)/Ta(9)/Nb ($d_1$)/Co$_{20}$Fe$_{60}$B$_{20}$(1)/MgO(0.9-3.5)/ Co$_{20}$Fe$_{60}$B$_{20}$(0.9-1.4)/Nb ($d_2$) /Ta(5)/Ru(15), where numbers in parentheses are thicknesses in nm and hyphenated numbers indicate a linearly varying thickness in certain layers across the specimen. pMTJs with different Nb thicknesses $d_1$ and $d_2$ have been fabricated for different purposes as detailed below. Circular junctions with diameters ranging from 100 nm to 100 μm were patterned and subsequently annealed under varying conditions. Detailed information on sample fabrication and characterization can be found in our previous publications.[39–41]

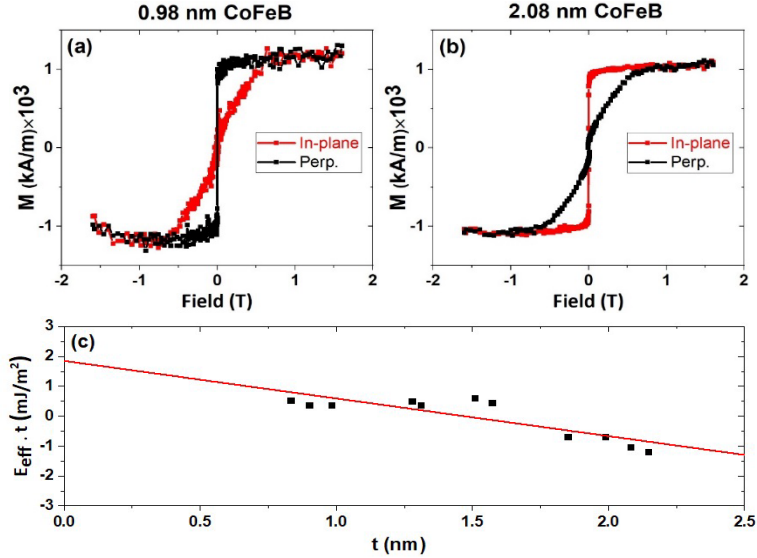

FIG. 1. (a) and (b) Hysteresis loops of the samples with CoFeB thicknesses of 0.98 and 2.08 nm, respectively. The samples were annealed at 300 °C for 10 min. (c) The plot of $E_{eff} \cdot t$ as a function of the CoFeB thickness t in a series of samples.

First, the magnetic properties of the films were studied by vibrating sample magnetometry (VSM) as shown in Figure 1. To isolate the role of the Nb buffer layer on the magnetic properties of CoFeB, an additional stack structure was produced, consisting of Ta(3)/Ru(5)/Ta(3)/Nb(10)/CoFeB(0.8-2.2)/MgO (2)/Ta(10)/Ru(20), where numbers in parentheses are thicknesses in nm. Figure 1(a) shows the easy axis along the out-of-plane direction for the sample with 0.98 nm CoFeB. The saturation magnetization ($M_s$) and the anisotropy field ($\mu_0 H_k$) of this sample was determined to be 1162 kA/m (1162 emu/cm³) and 0.67 T, respectively. The easy axis changes to the in-plane direction when the top CoFeB layer thicker than 1.85 nm and the hysteresis loops of the sample with a 2.08 nm top CoFeB is shown in Figure 1(b). For each sample, the effective perpendicular magnetic anisotropy energy density ($E_{eff}$) can be obtained as $E_{eff} = \frac{1}{2} M_s \cdot H_k$. The interfacial PMA energy density ($E_i$) can be obtained by $E_{eff} = E_b - 2\pi M_s^2 + \frac{E_i}{t}$, where $E_b$ and $2\pi M_s^2$ are bulk anisotropy energy density and shape anisotropy density, respectively, and $t$ is the thickness of CoFeB on top of MgO. In figure 1(c), $E_{eff} \cdot t$ is plotted as a function of $t$, and the y-intercept gives $E_i$, which is found to be 1.85 mJ/m² (1.85 erg/cm²) in this study for the samples annealed at 300 °C for 10 min. Unfortunately, PMA could not be maintained when the films were annealed at 400 °C. More research is needed as stability at 400 °C is required by the back-end-of-line integration of MTJ with CMOS. It was widely recognized that the formation energy between the HM layer and Fe (60% Fe in the CoFeB alloy used here) plays an important role in determining the thermal stability of pMTJs during annealing. [23,26] Generally, larger formation energy indicates the HM layer and Fe are less likely to diffuse into each other to form compounds. The formation energy of Nb-Fe is about -23 kJ/mol, which is substantially smaller than that of Mo-Fe (-3 kJ/mol), but it is similar to Ta-Fe (-22 kJ/mol).[42] From this point of view the thermal stability of Nb-pMTJs can be potentially increased to a degree that is similar to Ta-pMTJs through additional materials and process engineering.

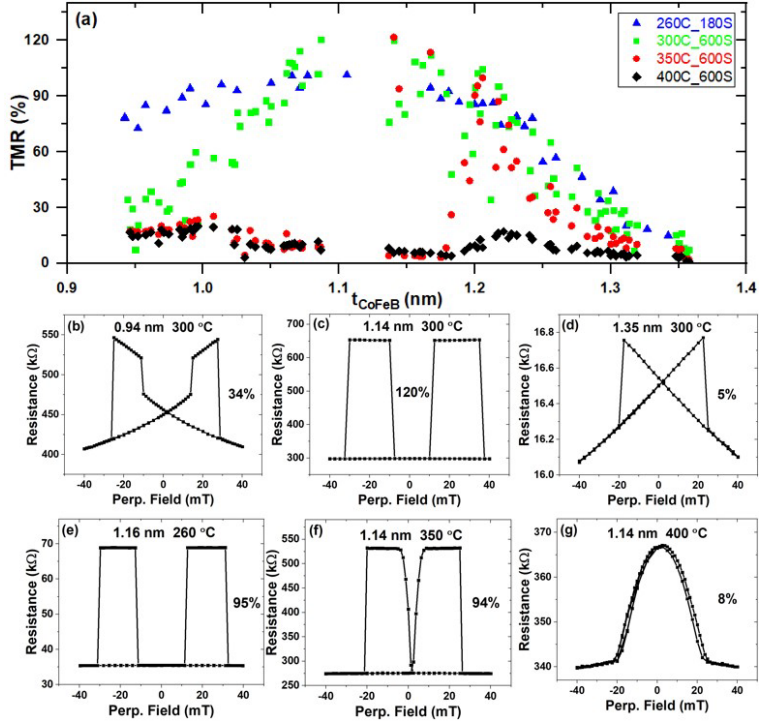

FIG. 2. (a) Evolution of TMR in Nb-pMTJs with different top CoFeB layer thicknesses after annealing under several different conditions. The samples were successively annealed at 260 °C for 3 min, 300 °C for 10 min, 350 °C for 10 min, and 400 °C for 10 min. (b-d) TMR hysteresis curves with (b) 0.94 nm, (c) 1.14 nm, (d) 1.35 nm top CoFeB thickness after the 300 °C annealing for 10 min. (e-g) TMR curves under different annealing conditions with similar top CoFeB thicknesses (around 1.14 nm).

Next, the TMR of the Nb-pMTJ was investigated in a series of samples with Nb thickness $d_1 = d_2 = 10$ nm and the top CoFeB layer thickness ranging from 0.9 nm to 1.4 nm. Fig. 2(a) shows the thickness dependence of the TMR of these junctions after different annealings. After the 260 °C annealing, the TMR values (blue triangles) almost stayed the same (about 90 %) when the thickness of the top CoFeB layer varies from 0.95 nm to 1.25 nm. After 300 °C annealing for 10 min (green squares), the TMR ratios were reduced in the junctions where the top CoFeB layer is thinner than 1.05 nm, as shown in Figure 2(b). This is likely due to a substantial magnetic dead layer thickness that has a more pronounced impact on the pMTJs with thinner CoFeB. The top CoFeB likely becomes superparamagnetic in this region. However, the maximum TMR increased to 120% with a 1.14 nm thick top CoFeB layer as shown in Figure 2(c), where a broad high-resistance plateau for the AP state and sharp switching for both the top and the bottom CoFeB layers can be seen. It is known that the thermal annealing is critical in determining the magnetoresistance[43] and defects level[44,45] in MTJs, and the increase of TMR is expected through the solid state epitaxy of CoFeB/MgO/CoFeB heterostructures.[46–49] But when the top CoFeB is too thick, such as 1.35 nm in this study (as shown in Fig. 2(d)), it exhibits in-plane anisotropy instead of PMA, which is the reason that there is a linear dependence of resistance for the top layer. The decrease of TMR (to nearly 15%) is more obvious in junctions with thinner top CoFeB after annealing at 350 °C (red dots). Finally, TMRs entirely collapsed after annealing at 400 °C as shown by the black diamonds. The evolution of TMRs for a pMTJ with 1.14 nm of top CoFeB is plotted in Figure 2(c), (f), (g), after the annealing at 300 ℃, 350 ℃ and 400 ℃, respectively. These figures can be examined with Figure 2(e), where the TMR curve for an MTJ with a very similar thickness of top CoFeB (1.16 nm) annealed at 260 ℃ is shown.  Good PMA for both top and bottom CoFeB layers can be seen after annealing at 260 ℃ and 300 ℃. The squareness of the perpendicular easy axis in the top CoFeB layer underwent a marked deterioration after annealing at 350 ℃, accompanied by a reduced TMR at 94%. Finally, the perpendicular magnetization of both CoFeB layers was lost after the annealing at 400 ℃ and the TMR further dropped to 8%.

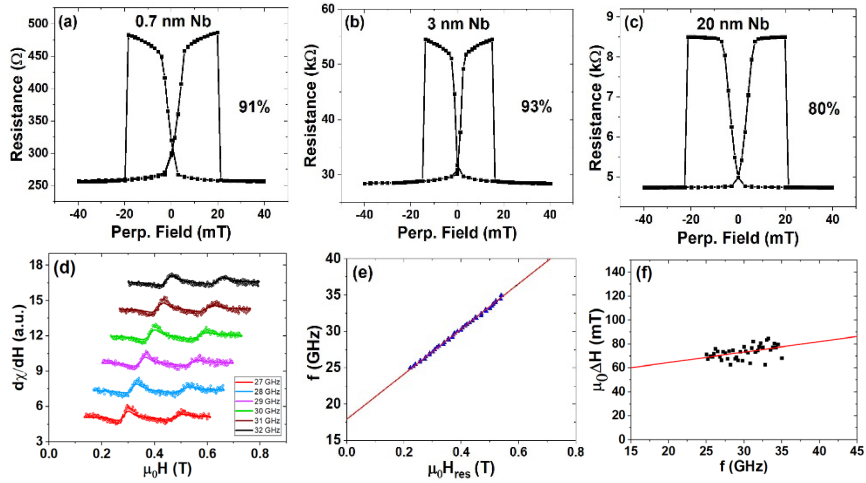

FIG. 3. The TMRs of a series of pMTJs with (a) $d_1$ = 0.7 nm, $d_2$ = 1.18 nm; (b) $d_1$ = 3 nm, $d_2$ = 3 nm; (c) $d_1$ = 20 nm, $d_2$ = 10 nm; after being annealed at 300 ℃ for 10 min. (d) Ferromagnetic resonance (FMR) measurements in Nb-pMTJs with FMR frequencies between 27 GHz and 32 GHz. (e) The fitting of the FMR frequency $f$ versus frequency-dependent resonance field $\mu_0 H_{res}$. (f) The fitting of the linewidth $\mu_0 \Delta H$ versus the FMR frequency $f$.

It was discovered previously that a thin dusting layer of Mo inserted between the Ta and CoFeB layers could give rise to a TMR ratio that is higher than using thick Ta or Mo layers. The TMR enhancement was attributed to the improved thermal stability brought by the Mo dusting layer (<1 nm), and the unique feature of Ta as an effective Boron sink.[23] Here a series of pMTJs with different Nb layer thicknesses were fabricated and the TMR was investigated after annealing at 300 ℃ for 10 min. Relatively small TMR ratios, 91%, and 93% were found in the junctions with 0.7 nm and 3 nm of buffer Nb as shown in Figure 3(a) and 3(b), respectively.  The low TMR is related to the lack of a good AP plateau, due to the weak PMA of the top CoFeB layers. The TMR ratios of these junctions are smaller than that of the 120% observed in the pMTJ with 10 nm of buffer Nb as shown in Figure 2(c). However, further increasing the thickness of Nb beyond 10 nm turned out to be deleterious – the TMR ratio is reduced to 80% with 20 nm buffer Nb. Previously a superconducting critical temperature of about 6 K was measured in an 8 nm-thick Nb film. [50] Therefore Nb films with the thickness of 10 nm and 20 nm in our structures are likely to be superconducting as well at low temperatures.

The spin dynamics of the samples were evaluated by broadband ferromagnetic resonance (FMR) measurements in Nb-pMTJs with $d_1$ = 20 nm and $d_2$ =10 nm. To increase the FMR sensitivity, an eternal field $H_{ext}$ was modulated ($\mu_0 H_{mod}$ = 1 mT peak amplitude) and a lock-in detection scheme was used, whereby FMR spectra were measured at fixed microwave frequencies (20 GHz - 40 GHz) under a swept $H_{ext}$. Fig. 3(d) shows illustrative FMR measurements for perpendicular $H_{ext}$, from which we extract the frequency-dependent resonance field $H_{res}$, and the linewidth $\Delta H$. The solid lines reflect the best fit of the raw absorption data. The ferromagnetic resonance relationship between microwave excitation frequency and resonant perpendicular applied magnetic field is given by: $\left(\frac{f}{\gamma}\right)_\perp = \mu_0 H_{res} + \mu_0 H_{eff}^\perp$, and linewidth versus frequency dispersion are given by: $\mu_0 \Delta H = \frac{2\alpha}{\gamma} f + \mu_0 \Delta H_0$, where $f$ is the FMR frequency, $\gamma$ is the gyromagnetic ratio, $\mu_0$ is the vacuum permeability, $\alpha$ is the Gilbert damping parameter and $\mu_0 H_{eff}^\perp = \mu_0 H_K - \mu_0 M_s$ is the effective perpendicular anisotropy field. FMR signals from both the top and bottom CoFeB layers of the pMTJ are shown in Figure 3(d). The effective anisotropy field was determined to be 0.58 T for the sample annealed at 260 °C for 3 min. From the fitting of $f$ versus frequency $\mu_0 H_{res}$ (Figure 3(e)) and the fitting of $\mu_0 \Delta H$ versus frequency $f$ (Figure 3(f)), the Gilbert damping factor α for the bottom CoFeB layer is extracted to be 0.011 for the sample annealed at 260 °C, which increased slightly to 0.013 after the annealing at 300 °C. These α values are similar to those found in W/CoFeB/MgO and Ta/CoFeB/MgO structures.[51]

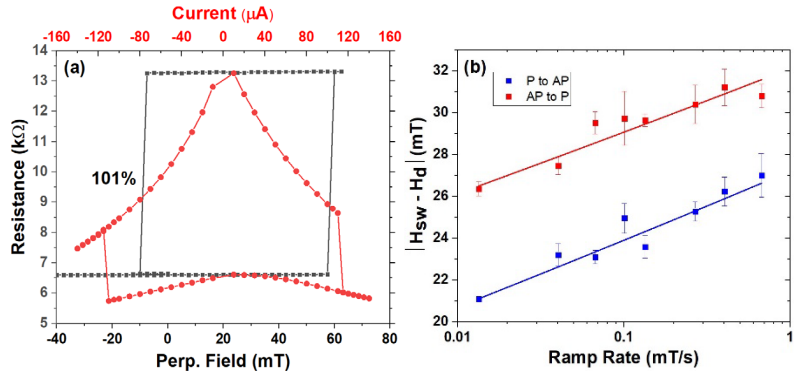

FIG. 4. (a) The TMR curve of the 145 nm-diameter MTJ nanopillar with 101 % TMR ratio (black squares) and the current-induced resistance switching (red dots) by the STT effect. (b) The ramp-rate-dependent switching field measurements for both P-to-AP and AP-to-P transitions.

Finally, we investigated the STT effect in the Nb-pMTJs. The core structure of these pMTJs for studying STT is Ta(8)/Ru(7)/Ta(9)/Mo(1.2)/CoFeB(1)/MgO (0.9)/CoFeB(1.2-1.7)/Nb(10)/Ta(5)/Ru(7), where numbers in parentheses are thicknesses in nm. Since the free layer is the CoFeB above the MgO, $d_1$ = 0, and $d_2$ = 10 nm were employed in these junctions to study the impact of Nb adjacent to the free layer. The magnetoresistance curve of a 145 nm device is plotted as black squares in Fig. 4(a), where the TMR ratio is 101 % with a resistance area product of 107 $\Omega \cdot \mu m^2$. The current-induced switching of resistance is plotted by the red dots in the same figure. The pMTJ can be fully switched between the *P* and *AP* states by the STT effect and the switching current is around 120 $\mu$A, corresponding to a quasistatic switching current density of $7.3 \times 10^5$ A/cm$^2$. An external field of 24 mT was applied to offset the dipolar field during the current-induced switching experiment. To determine the thermal stability of the pMTJ, we measured the TMR curves under different ramp rates of the magnetic field. The switching field should take the form[52]

$$<H_c> = H_{c0}(T)\left\{1 - \left[\frac{k_B T}{E_a}\ln\left(\frac{1}{\tau_0|R_H|ln2}\right)\right]^{\frac{2}{3}}\right\}, \quad (1)$$

where $k_B$ is Boltzmann's constant, $R_H$ is the ramp rate for the field, $H_{sw}$ is the switching field, and $H_d$ is the dipole field from the reference layer. From the fits to the field-ramp data in Fig. 4(d), we obtain E$_{P-AP}$ = 1 eV (corresponding to the thermal stability factor Δ = 40) for P-to-AP switching, and E$_{AP-P}$ = 1.18 eV (corresponding to Δ = 47) for AP-to-P switching.

In summary, we have investigated the magnetic, transport, and spin dynamic properties of the MTJs with Nb as the HM layers. Our results here demonstrate the compatibility of Nb with high-quality pMTJs, as reflected by the sizable TMR, the reasonably small damping parameter, and STT switching current density. These findings may be useful for future quantum computing and cryogenic memory applications.

**Acknowledgments**

This work was supported in part by Semiconductor Research Corporation through the Logic and Memory Devices program, by DARPA through the ERI program (FRANC), and by NSF through DMR-1905783. A.E. and K.W. were supported by the REU supplement of NSF ECCS-1554011.

**Data availability**

The data that support the findings of this study are available from the corresponding author upon reasonable request.